\documentclass[12pt]{article}

\newcommand{\bbr}{I\!\! R}
\newcommand{\bbz}{Z\!\!\! Z}

\newcommand{\x}{arXiv:}
\newcommand{\m}{\mathrm}

\usepackage{graphicx}

\begin{document}
\thispagestyle{empty}
\begin{center}

\null \vskip-1truecm \vskip2truecm {\Large{\bf

\textsf{The Geometry of The Entropic Principle and the Shape of the
Universe}

}}

\vskip1truecm {\large \textsf{Brett McInnes}} \vskip1truecm

 \textsf{National
  University of Singapore}

email: matmcinn@nus.edu.sg\\

\end{center}
\vskip1truecm \centerline{\textsf{ABSTRACT}} \baselineskip=15pt
\medskip
Ooguri, Vafa, and Verlinde have outlined an approach to
two-dimensional accelerating string cosmology which is based on
topological string theory, the ultimate objective being to develop
a string-theoretic understanding of ``creating the Universe from
nothing". The key technical idea here is to assign \emph{two
different} Lorentzian spacetimes to a certain Euclidean space.
Here we give a simple framework which allows this to be done in a
systematic way. This framework can be extended to higher
dimensions. We find then that the general shape of the spatial
sections of the newly created Universe is constrained by the OVV
formalism: the sections have to be flat and compact.

 \vskip3.5truecm
\begin{center}

\end{center}

\newpage

\addtocounter{section}{1}
\section* {\large{\textsf{1. The Cosmology of Topological Strings}}}
Ooguri, Vafa, and Verlinde have put forward \cite{kn:OVV} [see also
\cite{kn:dijk}] a novel approach to quantum cosmology, one in which
the topological string partition function is related to the wave
function of a two-dimensional Universe \cite{kn:hartlehawk} in a
mini-superspace description. One thus obtains a ``Hartle-Hawking
wave function for flux compactifications". \emph{This opens up the
prospect of a string-theoretic account of the creation of the
Universe from ``nothing"} \cite{kn:vilenkin1}\cite{kn:vilenkin2}.
The hope is that, if these ideas can be made to work in four
dimensions, it will be possible to construct a much-needed string
vacuum selection principle\footnote{See
\cite{kn:tye}\cite{kn:sarangi}\cite{kn:sashtye}\cite{kn:laura1}.}.
In particular, Ooguri et al are interested in using their wave
function to constrain the initial \emph{geometry} of the newly
created Universe\footnote{Another attempt to achieve this goal,
based instead on the ideas of \cite{kn:tye}, is described in
\cite{kn:tallandthin}.}.

Even at the classical or semi-classical levels, the techniques of
Ooguri et al have some very unusual features. In this note we
explain the geometric meaning of one of the crucial innovations in
\cite{kn:OVV}, and use this explanation as a guide to what might
be involved in extending these methods to Universes with
dimensions higher than two. We also point out that, even without a
detailed understanding of the OVV wave function in the
higher-dimensional case, we can use self-consistency conditions to
draw conclusions about the nature of the initial geometry
predicted by that wave function.

Ooguri et al \cite{kn:OVV} work with a compactification of IIB
string theory to Euclidean two-dimensional anti-de Sitter space,
the hyperbolic space H$^2$ of curvature $-$1/L$^2$; they use the
foliation of H$^{\m{n}}$, familiar from studies of the AdS/CFT
correspondence, by \emph{flat} slices \cite{kn:MAGOO}. However ---
and this is crucial for the cosmological application --- they then
perform a further compactification to a space with geometry
H$^2$/$\bbz$ and topology $\bbr\,\times\,\m{S}^1$, where S$^1$ is
a circle. The metric is
\begin{equation}\label{eq:P}
g(\m{H}^2/\bbz)_{++}\;=\;\m{K^2\,e^{(2\,\rho/L)}\,d\tau^2\;+\;d\rho^2};
\end{equation}
here, $\tau$ is an angular coordinate on a circle with radius K at
$\rho$ = 0. Notice that this can easily be generalized to a simple
``partial compactification" of H$^{\m{n+1}}$ for any n, of the form
H$^{\m{n+1}}$/$\bbz^{\m{n}}$, with topology $\bbr\,\times\,\m{T^n}$,
where $\m{T^n}$ is the n-torus.

The crux of the Ooguri et al construction is the remarkable claim
that this Euclidean space can be interpreted in \emph{two different
ways}, as follows. From the point of view of the topological string
it corresponds to a local anti-de Sitter geometry; $\tau$ is
``Euclidean time", leading to a Witten index which counts the
degeneracy of ground states of a certain black hole configuration.
But Ooguri et al \emph{also} interpret this geometry as a sort of
Euclidean version of de Sitter spacetime. The idea here is that
$\rho$ is \emph{equally entitled} to be regarded as some kind of
``Euclidean time" --- \emph{there is no way to distinguish one
coordinate as ``time"} on a Euclidean manifold. The manifold with
metric given in equation (\ref{eq:P}), with its circular [toral]
sections which ``expand" exponentially as $\rho$ increases, then
somehow corresponds to a Lorentzian accelerating universe, and this
is how the topological string partition function makes contact with
cosmology.

Of course, as Ooguri et al themselves point out, this second
interpretation cannot be taken literally within the usual
Hartle-Hawking formalism, since complexifying $\rho$ in the usual
manner of Euclidean quantum gravity would \emph{complexify the
metric itself}; so the ``correspondence" of equation (\ref{eq:P})
with de Sitter spacetime is obscure.

Nevertheless the idea that H$^2$/$\bbz$ could have a \emph{double
interpretation} is the key device which, in \cite{kn:OVV}, is
supposed to implement the connection between the topological
string and the wave function of the Universe. Furthermore, the
basic proposal that both $\rho$ and $\tau$ are equally entitled to
be regarded as ``Euclidean time" is clearly reasonable. We see
that, in order for the Ooguri et al programme to proceed, we have
to answer the first of two basic questions: \emph{how can we make
sense of the intuition that the metric in equation (\ref{eq:P})
somehow corresponds to an accelerating cosmology?} We can
formulate this question more concretely as: how is it possible for
there to be two distinct \emph{complexifications} [one AdS-like,
the other dS-like] of H$^2$/$\bbz$?

In the course of investigating this question, we soon realize that
the two-dimensional situation considered by Ooguri et al is very
special, because in that dimension the symmetry group of
(n+1)-dimensional anti-de Sitter spacetime, O(2,n), is isomorphic to
the symmetry group of (n+1)-dimensional de Sitter spacetime,
O(n+1,1). From this point of view, the double interpretation of
H$^2$/$\bbz$ \emph{must} be valid in some sense in the
two-dimensional case. For if the de Sitter group is the same as the
anti-de Sitter group, and granted that H$^2$ can be analytically
continued to AdS$_2$, then there \emph{must} be some natural way of
associating [some version of] H$^2$ with dS$_2$. But O(2,n) is
certainly not isomorphic to O(n+1,1) for higher n, and this
immediately raises our second question: does the double
interpretation of H$^2$/$\bbz$ \emph{only} work in two dimensions?

Our tactic for dealing with these questions focuses on a simple
ambiguity in the procedure of \emph{complexification}. The ambiguity
arises partly from the apparently trivial observation that, at least
for orientable spacetimes, the ($-\;+\;+\;+$) signature is in no way
preferable to ($+\;-\;-\;-$) signature, and partly from the
observation of Ooguri et al that there is no unique way of deciding
how to define Euclidean ``time".

Because of its traditional association with the ``no-boundary"
proposal in Euclidean quantum gravity, the process of
complexification is usually held to lead \emph{uniquely} from the
four-sphere to [global] de Sitter spacetime. We begin with a
critical review of this idea. We show that there is a sense in which
a \emph{local} anti-de Sitter metric can also be obtained by
complexifying the sphere. However, this construction leads to a
spacetime in which the spatial sections immediately contract after
the universe is created, and so the claim that complexification
leads uniquely from the sphere to a spacetime \emph{in which the
spatial sections eventually reach macroscopic size} can be
justified. The ideas of Ooguri et al prompt us to investigate
whether an analogous claim is justified in the case of hyperbolic
space: does complexification lead uniquely to AdS? The answer is no:
one can obtain [several physically acceptable versions of] de Sitter
spacetime in this way. Using these ideas, we give a concrete
interpretation of the double interpretation of H$^2$/$\bbz$ needed
for the arguments of Ooguri et al.

When we attempt to extend the construction to higher dimensions,
we find that it only makes sense for cosmological models with flat
[toral] spatial sections. Thus the \emph{shape} of the Universe is
``emergent" in this formalism, in precisely the same way [as has
recently been argued by Hartle \cite{kn:hartle}] that Lorentzian
signature emerges from the Hartle-Hawking wave function. It
remains to be seen whether the initial \emph{size} likewise
``emerges" from the OVV wave function.

\addtocounter{section}{1}
\section*{\large{\textsf{2. Complexification Ambiguity: The Case of the Sphere}}}
Complexification is a way of assigning a Lorentzian manifold to a
Euclidean one. It has been applied to curved spacetimes in at
least \emph{two distinct ways}.

First, it appeared as a way of studying the remote past of de Sitter
spacetime: in the Hartle-Hawking approach, one uses complexification
to convert the contracting half of de Sitter spacetime to half of a
four-sphere S$^4$, which remains attached to the [still Lorentzian]
upper half of dS$_4$. [Here we refer to ``SSdS", the Spatially
Spherical version of de Sitter spacetime; see below.]

The second major application of complexification was to string
theory, in the form of the AdS/CFT correspondence
\cite{kn:MAGOO}\cite{kn:wittenads}. Usually ``AdS" here means H$^4$,
the hyperbolic space, from which AdS$_4$ can be obtained in this
way. The work of Ooguri et al essentially constructs the analogue of
the Hartle-Hawking wave function, defined on a hyperbolic space
instead of a sphere. We therefore need to ask precisely how
complexification works for hyperbolic space. As a preparation for
that, let us review the spherical case in the light of the
observations made in \cite{kn:OVV}.

It is a basic fact that if one takes the standard metric on the
four-sphere of radius L,
\begin{eqnarray}\label{eq:D}
g(\mathrm{S^4})_{++++}\; =\;
\m{L}^2\,\Big\{\m{d\xi^2}\;+\;\m{cos^2(\xi)}\,[\mathrm{d\chi^2\; +\;
sin^2(\chi)}\{\mathrm{d}\theta^2 \;+\;
\mathrm{sin}^2(\theta)\,\mathrm{d}\phi^2\}]\Big\},
\end{eqnarray}
where all of the coordinates are angular, and continues
$\xi\,\rightarrow\,\m{iT/L}$, then the result is the \emph{global}
de Sitter metric, in its Spatially Spherical form,
\begin{eqnarray}\label{eq:E}
g(\mathrm{SSdS_4})_{-\,+++}\; =\;-\;
\m{dT}^2\;+\;\m{L}^2\,\m{cosh^2(T/L)}\,[\mathrm{d\chi^2\; +\;
sin^2(\chi)}\{\mathrm{d}\theta^2 \;+\;
\mathrm{sin}^2(\theta)\,\mathrm{d}\phi^2\}],
\end{eqnarray}
with the indicated signature.

Motivated by the arguments of Ooguri et al \cite{kn:OVV}, we now
observe that $\xi$ and $\chi$ have the same status as angles, and
the same range; \emph{there is no justification for preferring one
to the other}. Let us complexify $\chi$ in equation (\ref{eq:D})
instead of $\xi$, replacing $\chi\,\rightarrow\,\m{\pm is/L}$, and
for convenience relabelling $\xi$ as u/L [without complexifying it].
We obtain, since sin$^2(\chi)$ reverses sign under complexification
of $\chi$,
\begin{eqnarray}\label{eq:F}
g(\mathrm{DAdS_4})_{+---}\; =\;
\m{du^2}\;-\;\m{cos^2(u/L)}\,[\mathrm{ds^2\; +\;\m{L^2}\,
sinh^2(s/L)}\{\mathrm{d}\theta^2 \;+\;
\mathrm{sin}^2(\theta)\,\mathrm{d}\phi^2\}].
\end{eqnarray}
But, \emph{purely locally}, this is the \emph{anti}-de Sitter
metric, in ($+\;-\;-\;-$) signature, and expressed in terms of
coordinates \cite{kn:hawking}\cite{kn:gibbons}\cite{kn:orbifold}
based on the timelike geodesics which are perpendicular to the
spatial sections; the coordinate u is proper time along these
geodesics. These coordinates do not cover the entire spacetime, of
course, because these timelike geodesics intersect, being drawn
together by the attractive nature of gravity in anti-de Sitter
spacetime [which satisfies the Strong Energy Condition]. This is why
these coordinates give the false impression that there is no
timelike Killing vector in AdS --- there is one, but it does not
correspond to the time coordinate u. On the other hand, these
coordinates do have the virtue of reflecting the behaviour of
inertial observers in AdS$_4$. In fact, these coordinates cover the
Cauchy development of a single spacelike slice: in this sense they
are the precise analogues of the standard global coordinates in
dS$_4$, which happen to cover the entire spacetime simply because
gravity is repulsive in that case, ensuring that the worldlines of
inertial observers do not intersect.

The fact nevertheless remains that in continuing (\ref{eq:D}) to
(\ref{eq:F}) we have only continued S$^4$ to a small \emph{part} of
AdS$_4$; see \cite{kn:braga} for a detailed discussion of related
issues. This part is of course extensible; objects can leave or
enter the spacetime without encountering any singularity. This does
not make sense physically, \emph{particularly in the context of
``creation from nothing"}; for if objects or signals can enter the
spacetime along a null surface, it is doubtful that one can claim
that the Universe was ``created" on a specific spacelike surface.

Fortunately there is an extremely natural way to solve this
problem, as follows. The Lorentzian metric (\ref{eq:F}) is of
course a purely local structure. We are not told how to select the
global structure from the large range of possibilities compatible
with this local metric. In particular, the spatial sections here
have the geometry of three-dimensional hyperbolic space. These can
be compactified: that is, we interpret the spatial part of the
metric as a metric on a space of the form H$^3/\Gamma$, where
$\Gamma$ is some discrete freely acting infinite group of H$^3$
isometries such that the quotient is compact. The compactified
spacetime is incomplete only in the past, not along null surfaces,
and so it makes sense to speak of it being created along a
spacelike surface. We shall discuss this in more detail below. For
the present we merely note that, with this interpretation,
\emph{one loses the global timelike Killing vector} defined on
full AdS$_4$: it does not project to a Killing vector on the
quotient. The metric in (\ref{eq:F}) is a genuinely \emph{dynamic}
metric on a spacetime, with topology
$\bbr\,\times\,(\bbr^3/\Gamma)$, with spatial sections which
expand from zero size and then contract back to zero
size\footnote{The initial and final points do not correspond to
curvature singularities, but generically they would: this geometry
fails to satisfy the \emph{generic condition} given on page 266 of
\cite{kn:hawking}. The singularity theorems imply that the
slightest generic perturbation causes these points to become
genuinely singular.}. We can call this ``Dynamic AdS$_4$" [hence
the notation in (\ref{eq:F})]; \emph{the timelike symmetry has
been broken topologically}. It follows that this spacetime is
physically distinct from true AdS$_4$.

Thus, the alternative complexification of S$^4$ does not lead to
AdS$_4$, but rather to a spacetime with compact spatial sections
with a non-trivial evolution controlled by the metric (\ref{eq:F}).
If this spacetime is ``created from nothing" along its spacelike
surface of zero extrinsic curvature at u = 0, then it will
immediately begin to contract; thus it will never reach macroscopic
size.

We conclude that there is a ``complexification ambiguity" for the
sphere, in the sense that the sphere can indeed be continued to two
distinct Lorentzian spacetimes. But one of these continuations fails
to attain macroscopic size, so we have a concrete justification for
discarding it. [This argument is modelled on Hartle's
\cite{kn:hartle} discussion of the ``emergence" of Lorentzian
signature; see below.] In this sense, we can continue to claim that
there is only one physically significant continuation of the sphere.

Nevertheless, there are some interesting lessons here. First, it
is clear more generally that Euclidean spaces will sometimes have
more than one Lorentzian version if we accept both ($+\;-\;-\;-$)
and ($-\;+\;+\;+$) signatures. Actually, for the most general
topologically non-trivial spacetimes, the two possible choices of
signature \emph{are not fully physically equivalent}, a surprising
fact first pointed out by Carlip and DeWitt-Morette
\cite{kn:carlip}. This shows that the distinction we are
discussing here is by no means trivial. Nevertheless, none of the
issues raised in \cite{kn:carlip} actually arise here --- all of
our spacetimes, including the topologically non-trivial ones like
Dynamical AdS, are orientable in all of the possible senses ---
so, in our case, there can be no initial justification for
preferring one signature to the other. We merely propose to take
this observation seriously when applying complexification.

A second lesson to be drawn from the discussion above is that one
should bear in mind that \emph{the sign of the curvature of a
Lorentzian manifold depends on the convention for signature.} Thus
anti-de Sitter spacetime is a spacetime of \emph{positive} curvature
in ($+\;-\;-\;-$) signature, while de Sitter spacetime has
\emph{negative} curvature in that convention. Hence the association
of a \emph{positively} curved version of AdS$_4$ with the sphere is
perhaps not so surprising. This of course opens the way to
justifying the hope of Ooguri et al, that some version of de Sitter
spacetime can emerge from their basic negatively curved Euclidean
space.

In this section we have seen that one has to qualify the claim that
de Sitter spacetime is \emph{the only} Lorentzian continuation of
the four-sphere: one can also obtain a \emph{local} version of
anti-de Sitter spacetime in this way. Thus complexification is
``locally ambiguous" in this sense. However, \emph{in the case of
the sphere,} global considerations effectively remove the ambiguity,
since we saw that one certainly cannot obtain \emph{global} anti-de
Sitter spacetime by complexifying the sphere: instead one obtains a
spacetime which is no sooner created than it shrinks to
non-existence. In this sense, the usual understanding is correct: de
Sitter spacetime is indeed the unique \emph{macroscopic}
complexification of the sphere.

The situation becomes more interesting in the case of hyperbolic
geometry, however, as we shall now see.

\addtocounter{section}{1}
\section*{\large{\textsf{3. Complexification Ambiguity: The Case of Hyperbolic Space}}}
The hyperbolic Euclidean space H$^4$, with its metric of constant
curvature equal to $-$1/L$^2$, can be defined as a connected
component of the locus

\begin{equation}\label{eq:G}
\m{- \,A^2\;+\: B^2\; + \;X^2 \;+ \; Y^2\; +\; Z^2\; =\; -\,L^2},
\end{equation}
defined in a five-dimensional Minkowski space. It is clear that all
of the coordinates except A can range in ($-\,\infty,\;+\,\infty$),
while A has to satisfy $\m{A^2\;\geq\;L^2}$. We always pick the
connected component on which A is positive.

Hyperbolic space can be globally foliated in a variety of
interesting ways. The leaves of the foliation are distinguished by a
parameter; \emph{this parameter plays the role of Euclidean time}.
Clearly there is no ``preferred" foliation. This statement is the
generalized version of the observation, made by Ooguri et al, that
both $\rho$ and $\tau$ are equally entitled to be interpreted as
``time" in equation (\ref{eq:P}).

We shall now discuss four physically interesting ways of foliating
H$^4$, and their various complexifications.

\subsubsection*{{\textsf{3.1. Foliation Corresponding to Anti-de Sitter}}}

\begin{figure}[!h]
\centering
\includegraphics[width=0.7\textwidth]{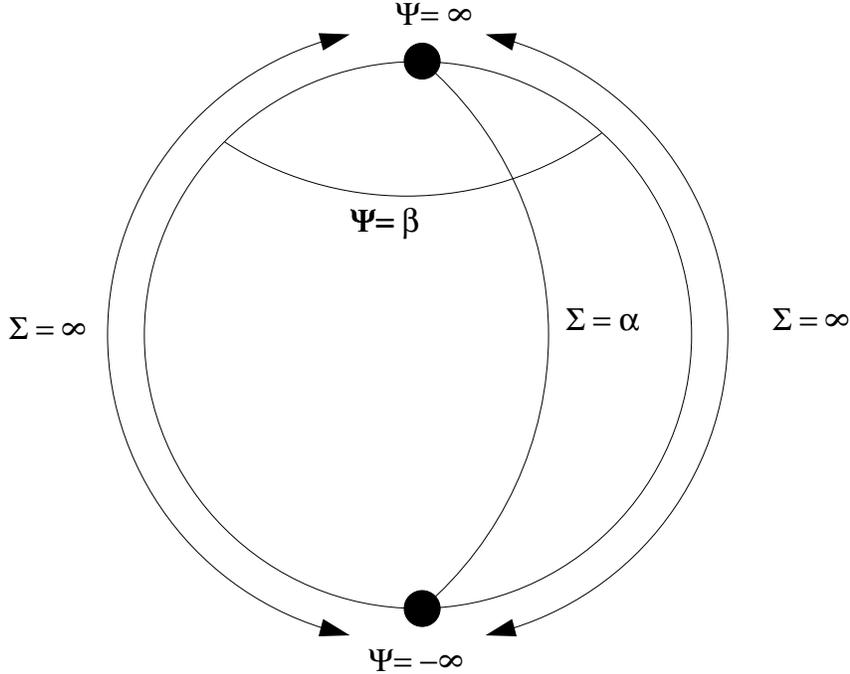}
\caption{Zero extrinsic curvature foliation of H$^4$.}
\end{figure}

The underlying manifold of H$^4$ can be regarded as the interior of
a ball, as shown in Figure 1; this is clear if we write equation
(\ref{eq:G}) as
\begin{equation}\label{eq:GORGONZOLA}
\m{B^2\; + \;X^2 \;+ \; Y^2\; +\; Z^2\; =\;A^2\; -\,L^2},
\end{equation}
since we clearly have a three-sphere at infinity. The conformal
boundary is also shown, but points on the boundary are of course not
points of H$^4$.

Choosing the connected component on which [in equation (\ref{eq:G})]
A is positive, we can pick coordinates
$\Psi$,$\Sigma$,$\theta$,$\phi$ such that
\begin{eqnarray} \label{eq:H}
\m{A} & = & \m{L\;cosh(\Psi)\;cosh(\Sigma)}                       \nonumber \\
\m{B} & = & \m{L\;sinh(\Psi)\;cosh(\Sigma)} \nonumber\\
\m{Z} & = & \m{L\;sinh(\Sigma)\;cos(\theta)}           \nonumber \\
\m{Y} & = & \m{L\;sinh(\Sigma)\;sin(\theta)\;cos(\phi)}  \nonumber \\
\m{X} & = & \m{L\;sinh(\Sigma)\;sin(\theta)\;sin(\phi)},
\end{eqnarray}
and these coordinates cover H$^4$ globally if we let $\Psi$ run from
$-\,\infty$ to $+\,\infty$ while $\Sigma$ runs from 0 to
$+\,\infty$. [Both $\theta$ and $\phi$ are suppressed in Figure 1.]

The surfaces $\Sigma$ = constant are topological cylinders; recall
that a cylinder, with topology $\bbr\times\m{S}^2$, is a
three-sphere from which two points have been deleted. If we think
of H$^4$ as the interior of a four-dimensional ball, then these
cylinders are ``pinched together" as they approach the boundary at
the points $\Psi\;=\;\pm\infty$. A typical ``pinched cylinder"
inside the boundary is shown\footnote{In dimensions above two, the
reader can picture such a cylinder by rotating the $\Sigma =
\alpha$ line about an axis passing through the heavy dots.} as
$\Sigma$ = $\alpha$ = constant in Figure 1. It is clear that the
conformal boundary is a cylinder with two additional points
[corresponding to $\Psi\;=\;\pm\infty$] added: with these
additions, the boundary becomes the familiar three-sphere. Notice
that the structure of the interior, partitioned into these
cylinders, is very similar to that of the simply connected version
of Lorentzian anti-de Sitter spacetime
\cite{kn:gibbons}\cite{kn:orbifold}. Notice too that the cylinders
themselves \emph{do not} intersect --- they only do so in the
conformal completion.

The foliation in which we are interested here is given by the
surfaces $\Psi$ = constant, transverse to the cylinders. These are
copies of three-dimensional hyperbolic space; all of them have
\emph{the same intrinsic curvature}, $-$1/L$^2$, as each other and
as the ambient H$^4$. This can be seen by noting that the first two
equations of (\ref{eq:H}) imply that $\m{-\,A^2\;+\;B^2}$ is
independent of $\Psi$. All of these slices intersect the conformal
boundary at right angles, and all of them have \emph{zero extrinsic
curvature}. A typical surface $\Psi$ = $\beta$ = constant is shown
in Figure 1.

Because the slices have zero extrinsic curvature, $\Psi$ can be
compactified: the consequences for Figure 1 are shown in Figure 2:
parts of the diagram have to be deleted, and the top and bottom of
the diagram have to be identified. The boundary changes topology
from S$^3$ to S$^1\,\times\,\m{S}^2$. This is one possible form of
``hyperbolic space" if, as is frequently the case, one wants
Euclidean time to be periodic.

\begin{figure}[!h]
\centering
\includegraphics[width=0.5\textwidth]{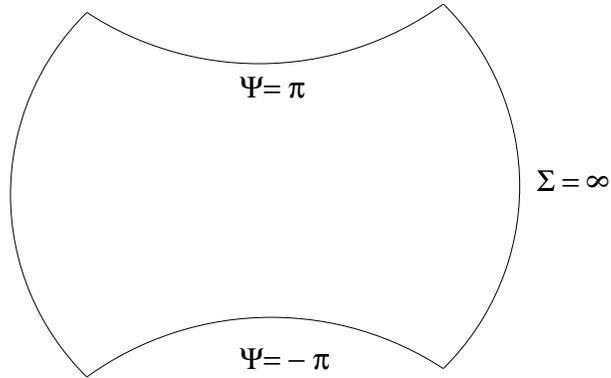}
\caption{ZEC foliation of H$^4$, with compactification of $\Psi$
axis.}
\end{figure}

The metric [with constant curvature $-$1/L$^2$] expressed in terms
of the coordinates defined by this Zero Extrinsic Curvature
foliation is
\begin{eqnarray}\label{eq:I}
g(\m{H^4}; \m{ZEC})_{++++}\;=\; \m{L^2\,\Big\{cosh^2(\Sigma) \,
d\Psi^2\;+ \;d\Sigma^2 \; +\; sinh^2(\Sigma)[d\theta^2 \; +\;
sin^2(\theta)d\phi^2]\Big\}}.
\end{eqnarray}

We now complexify $\Psi\,\rightarrow\,\m{iU/L}$ [keeping U/L
periodic if $\Psi$ is compactified as in Figure 2] and re-label
$\Sigma$ [without complexifying it] as S/L, then we obtain, from
(\ref{eq:I}),
\begin{eqnarray}\label{eq:L}
g(\m{AdS_4)_{-\,+++}\;=\; -\,cosh^2(S/L) \, dU^2\;+ \;dS^2 \; +\;
L^2\,sinh^2(S/L)[d\theta^2 \; +\; sin^2(\theta)d\phi^2]},
\end{eqnarray}
and this is precisely \cite{kn:hawking} the \emph{globally valid}
AdS$_4$ metric, in the indicated signature. The basic definition
of AdS, as a locus in a higher-dimensional space, leads to cyclic
time\footnote{Note that the hyperbolic functions of $\Psi$ in the
relations (\ref{eq:H}) become trigonometric, therefore periodic,
when $\Psi$ is complexified.}, so with angular U/L this is indeed
precisely AdS$_4$; while we are free to take the universal cover,
it can be argued \cite{kn:gibbons}\cite{kn:orbifold} that this is
not really necessary. In other words, Figure 2 is relevant in the
case where either Euclidean or Lorentzian time is periodic.

This calculation provides a rigorous basis for the standard claim
that anti-de Sitter spacetime is the complexified version of
hyperbolic space. It is clear, however, that we obtained this result
by \emph{choosing} a very specific foliation of H$^4$ --- one with
leaves having zero extrinsic curvature. By doing this, we guarantee
that the complexified version will have a timelike Killing vector;
but, again, this is a matter of deliberate construction, not
something that is forced on us. [Notice in this connection that if
we complexify $\Sigma$ instead of $\Psi$ in (\ref{eq:I}), then the
result is the \emph{static} version of the local de Sitter metric,
with ($+\;-\;-\;-$) signature.] If, following Ooguri et al, we
declare that other foliations of H$^4$ are acceptable, then we can
expect to obtain a dynamical spacetime upon complexification. Let us
see how this works.

\subsubsection*{{\textsf{3.2. Foliation Corresponding to Spatially Spherical de Sitter}}}
Now we shall consider a second, completely different, but also
entirely global foliation of H$^4$, shown in Figure 3.
\begin{figure}[!h]
\centering
\includegraphics[width=0.6\textwidth]{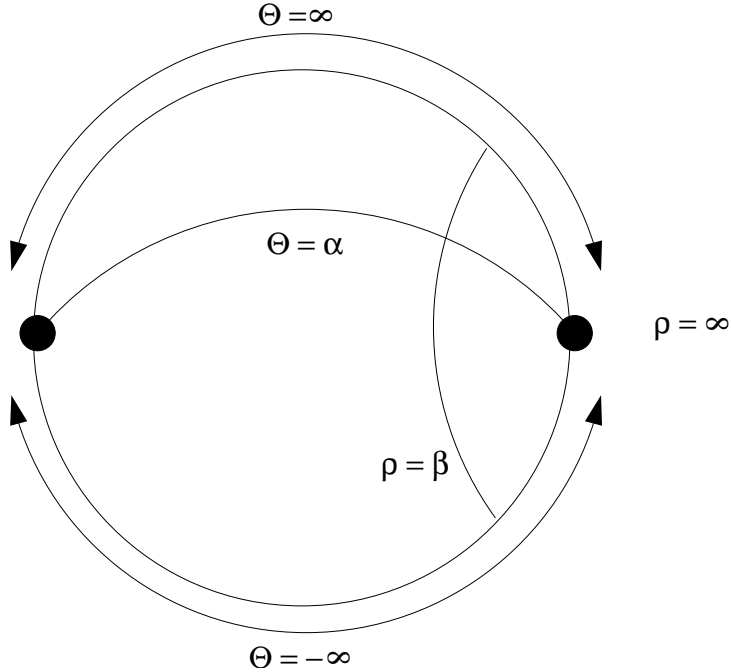}
\caption{$\Theta$ foliation of H$^4$.}
\end{figure}
Choose coordinates $\Theta$,$\rho$,$\theta$,$\phi$, where $\Theta$
runs from $-\,\infty$ to $+\,\infty$ while $\rho$ runs from 0 to
$+\,\infty$, and set
\begin{eqnarray} \label{eq:J}
\m{A} & = & \m{L\;cosh(\Theta)\;cosh(\rho)}                       \nonumber \\
\m{B} & = & \m{L\;sinh(\Theta)} \nonumber\\
\m{Z} & = & \m{L\;cosh(\Theta)\;sinh(\rho)\;cos(\theta)}           \nonumber \\
\m{Y} & = & \m{L\;cosh(\Theta)\;sinh(\rho)\;sin(\theta)\;cos(\phi)}  \nonumber \\
\m{X} & = & \m{L\;cosh(\Theta)\;sinh(\rho)\;sin(\theta)\;sin(\phi)}.
\end{eqnarray}
Then the leaves of the foliation are labelled by $\Theta$. We shall
call this the ``$\Theta$ foliation".

Because we are suppressing two angles, Figure 3 seems to resemble
Figure 1, but this is misleading [except in two dimensions, see
below]. Here the surfaces $\Theta$ = constant are, from the second
member of equations (\ref{eq:J}), just the submanifolds B =
constant; they are copies of the three-dimensional hyperbolic space
H$^3$, as can be seen at once from equation (\ref{eq:G}). This
foliation differs from the previous one in a crucial way, however:
whereas previously the slices all had the same intrinsic curvature,
$-1$/L$^2$, as the ambient space, here the surface $\Theta$ =
$\alpha$ = constant can be written as
\begin{equation}\label{eq:JACKAL}
\m{- \,A^2\;+\; X^2 \;+ \; Y^2\; +\; Z^2\; =\;
-\,L^2\,cosh^2(\alpha)},
\end{equation}
so the magnitude of the intrinsic curvature of a slice is reduced
by a factor of sech$^2$($\alpha$). The slices become flatter as
they expand towards the boundary. Their extrinsic curvature is
therefore \emph{never zero}, with the sole exception of the
equatorial slice at $\Theta$ = 0. In this case, the copies of
H$^3$ are all ``pinched together" as we move towards \emph{their}
boundaries, that is, as $\rho\,\rightarrow\,\infty$. A typical
H$^3$ slice, $\Theta$ = $\alpha$ = constant is shown in Figure 3.

Notice that the slices themselves do not intersect: only their
conformal completions do so. At any point actually in a given copy
of H$^3$ [and not on its conformal boundary], one can send a
geodesic [shown in Figure 3] of the form $\theta$ = $\phi$ =
constant, $\rho$ = constant = $\beta$, towards infinity, and this
will uniquely define two points on the boundary, one each at $\Theta
= \pm\infty$. From this point of view, one can say that the
conformal infinity of H$^4$ is ``finitely disconnected": the usual
boundary three-sphere is divided into two hemispheres corresponding
to the forward or backward ``evolution" along the geodesics
perpendicular to these slices. Of course, topologically the boundary
is connected, since the two hemispheres join along the common
conformal boundary of all of the slices; but this can only be
detected by proceeding to infinity in the ``spacelike" direction,
that is, along the slices.

It is clear that this foliation, like the previous one, foliates
H$^4$ globally, though it is in general totally different to the one
shown in Figure 1. The metric [with curvature $-$1/L$^2$] with
respect to this foliation is
\begin{eqnarray}\label{eq:K}
g(\mathrm{H}^4; \Theta)_{++++}\; =\;\m{L^2}\,\Big\{
\m{d\Theta^2}\;+\;\m{cosh^2(\Theta)}[\mathrm{d\rho^2}\; +\;
\mathrm{\,sinh^2(\rho)}\{\mathrm{d}\theta^2 \;+\;
\mathrm{sin}^2(\theta)\,\mathrm{d}\phi^2\}]\Big\}.
\end{eqnarray}

If we now complexify by mapping $\rho\,\rightarrow\,\pm \m{i\chi}$
while re-labelling $\Theta$ as T/L, we obtain, since sinh($\pm$
i$\,\chi$) = $\pm \m{i\;sin(\chi)}$,
\begin{eqnarray}\label{eq:M}
g(\m{SSdS_4)_{+---}\; =\; dT^2\;-\;L^2\,cosh^2(T/L)\,[d\chi^2\; +\;
sin^2(\chi)\{d\theta^2 \;+\; sin^2(\theta)\,d\phi^2\}]},
\end{eqnarray}
which is of course the global, Spatially Spherical form of the de
Sitter metric, but now in ($+\;-\;-\;-$) signature. We have become
accustomed to thinking of de Sitter spacetime as a space of
\emph{positive} curvature, but we again remind the reader that this
is a matter of convention [of the signature]: in ($+\;-\;-\;-$)
signature, de Sitter spacetime has \emph{negative} curvature, and
there is no sense in which this is less natural than positive
curvature in the opposite convention.

Thus de Sitter and anti-de Sitter spacetimes are seen to have a
common origin in \emph{different foliations} of the \emph{same}
Euclidean [hyperbolic] space\footnote{The idea that distinct
foliations can have distinct physics was proposed in
\cite{kn:tseytlin}\cite{kn:buchel}.}. In this case
--- unlike that of the sphere considered earlier
--- \emph{both} complexifications lead to the full, global
Lorentzian versions of the spacetimes in question.

Before proceeding, let us settle a technical point. One can regard
analytic continuation as a mere technical device, a
solution-generating technique. But if one wishes to use it in the
original way, to construct the Euclidean gravity path integral
\cite{kn:HG}, then it is important that the procedure should also
complexify the volume form. In equation (\ref{eq:K}), for example,
the volume form is
\begin{eqnarray}\label{eq:MONSTER}
\m{dV}(g(\mathrm{H}^4;\Theta)_{++++})\; =\;\m{L^4\,
cosh^3(\Theta)\,sinh^2(\rho)\,sin(\theta)\,d\Theta\,d\rho\,d\theta\,d\phi},
\end{eqnarray}
and one sees at once that complexifying $\rho$ [to obtain equation
(\ref{eq:M})] does indeed complexify the volume form. However, this
only works if the number of spacetime dimensions is \emph{even}.
Thus we shall confine ourselves to even spacetime dimensions
henceforth. [Depending on the dimension, one may have to choose the
sign of the imaginary factor in the complexification so that the
volume form ``rotates" in the correct direction. This is the reason
for the $\pm$ sign in the complexification of $\rho$, above.]

The observation that de Sitter spacetime, like anti-de Sitter
spacetime, has a natural association with hyperbolic space was
suggested in \cite{kn:exploring}; related ideas were investigated
in \cite{kn:bala}; it has been put on a rigorous mathematical
basis [though mainly in the case of Einstein bulks, which are of
limited cosmological interest] by Anderson \cite{kn:anderson}; and
it is relevant to any theory which makes use of the fact that the
de Sitter and anti-de Sitter spacetimes are mutually locally
conformal \cite{kn:eva}\cite{kn:triality}.

The fact that hyperbolic space can be complexified to de Sitter
spacetime gives reason to hope that it should indeed be possible to
realise the suggestion of Ooguri et al that the metric in equation
(\ref{eq:P}) has a cosmological interpretation. With this in mind,
we proceed to yet another foliation of hyperbolic space.

\subsubsection*{{\textsf{3.3. Foliation Corresponding to Spatially Hyperbolic de Sitter}}}

Since $\m{A\;\geq\;L}$ in equation (\ref{eq:G}), the most obvious
way to choose coordinates here is to define P,$\chi$,$\theta$,$\phi$
such that
\begin{eqnarray} \label{eq:HYPER}
\m{A} & = & \m{L\;cosh(P/L)}                       \nonumber \\
\m{B} & = & \m{L\;sinh(P/L)\;cos(\chi)} \nonumber\\
\m{Z} & = & \m{L\;sinh(P/L)\;sin(\chi)\;cos(\theta)}           \nonumber \\
\m{Y} & = & \m{L\;sinh(P/L)\;sin(\chi)\;sin(\theta)\;cos(\phi)}  \nonumber \\
\m{X} & = & \m{L\;sinh(P/L)\;sin(\chi)\;sin(\theta)\;sin(\phi)},
\end{eqnarray}
giving the familiar Poincar$\m{\acute{e}}$ ``disc" representation of
hyperbolic space, with Poincar$\m{\acute{e}}$ radial coordinate P
and metric
\begin{equation}\label{eq:ORNITHOPTER}
g(\mathrm{H^4; P})_{++++}\; =\;+\;
\m{dP}^2\;+\;\m{L^2}\,\m{sinh^2(P/L)\,[d\chi^2\; +\;
sin^2(\chi)}\{\mathrm{d}\theta^2 \;+\;
\mathrm{sin}^2(\theta)\,\mathrm{d}\phi^2\}].
\end{equation}
We are now foliating H$^4$ by spheres labelled by P, which as usual
is ``Euclidean time". In many ways this is the most natural way to
picture hyperbolic space
--- for example, it makes the [spherical] structure at infinity very clear.
However, this version of H$^4$ does not usually appear in
discussions of the Euclidean form of the AdS/CFT correspondence, for
the simple reason that the usual
($+\;+\;+\;+$)$\rightarrow$($-\;+\;+\;+$) continuation is not
possible here: in this case, \emph{only} the
($+\;+\;+\;+$)$\rightarrow$($+\;-\;-\;-$) continuation actually
works\footnote{By contrast, we saw that the metric in (\ref{eq:I})
has a second continuation to the static patch of de Sitter, and
similarly the metric in (\ref{eq:K}) can be continued also to
Dynamical AdS.}. For if we attempt to complexify P, the result is
not Lorentzian; while if we complexify $\chi$ and re-label suitably
we obtain
\begin{eqnarray}\label{eq:OLIPHAUNT}
g(\mathrm{SHdS_4})_{+\,---}\; =\;+\;
\m{d\tau}^2\;-\;\m{sinh^2(\tau/L)}\,[\mathrm{dr^2\; +\;\m{L^2}\,
sinh^2(r/L)}\{\mathrm{d}\theta^2 \;+\;
\mathrm{sin}^2(\theta)\,\mathrm{d}\phi^2\}].
\end{eqnarray}
This is actually yet another version of de Sitter spacetime, one
with \emph{negatively} curved spatial sections instead of local
spheres. This is the prototype for ``hyperbolic" accelerating
cosmologies, which have recently attracted much attention from
various points of view
\cite{kn:ohta}\cite{kn:susskind}\cite{kn:silver1}\cite{kn:lars}\cite{kn:silver2}.
This spacetime is therefore potentially of very considerable
interest.

As is explained in \cite{kn:leblond}, the coordinates in equation
(\ref{eq:OLIPHAUNT}) only cover the interior of the future
lightcone of a point on the equator of global de Sitter spacetime.
In this sense, this metric represents not a spacetime, but rather
a \emph{part} of a spacetime; this part is geodesically incomplete
in a way that is physically meaningless and that would forbid
``creation from nothing". As in our discussion of Dynamical AdS in
Section 2 above, we take this as an instruction to compactify the
spatial sections: we interpret the three-dimensional metric on the
spatial sections in equation (\ref{eq:OLIPHAUNT}) as a metric on a
compact space of the form H$^3/\Gamma$, where $\Gamma$ is some
discrete freely acting infinite group of H$^3$ isometries such
that the quotient is compact. One can think of this in the
following way: we are effectively imposing certain [very
intricate] restrictions on the ranges of the coordinates in the
spatial part of the metric\footnote{See \cite{kn:singularstable}
for a discussion of this in the much simpler case of Spatially
Toral de Sitter spacetime, STdS.}. Note that in general $\Gamma$
will be a very complicated object, but also that this complexity
may have a direct physical meaning in connection with the way
string theory may possibly resolve the ``singularity" at $\tau$ =
0; see \cite{kn:silver1}\cite{kn:silver2} for the details. In this
work, however, the geometry around $\tau$ = 0 will be regularized
in a different and more direct way, since in any case we will
truncate the spacetime at some non-zero value of $\tau$, in the
usual manner of ``creation from nothing".

\emph{Compactifying the spatial slices produces a physically
distinct spacetime.} It is no longer possible for objects to enter
the spacetime from ``outside", except at $\tau$ = 0. That is, the
spacetime is still geodesically incomplete, but only at $\tau$ =
0; this is reasonable physically, since the introduction of
conventional matter into this spacetime can in any case be
expected to generate a curvature singularity at that
point\footnote{Since the Strong Energy Condition does not hold
here, one cannot prove this using the classical singularity
theorems; instead one invokes the recent results of Andersson and
Galloway \cite{kn:andergall}; see also \cite{kn:gall} and
\cite{kn:singularstable} for a discussion.}. More relevantly here,
we are interested in creating these universes from ``nothing",
and, in that context, the spacetime would in any case be truncated
at some value of $\tau$ strictly greater than zero, where there
would be a transition from a Euclidean metric to a Lorentzian one;
so the incompleteness at $\tau$ = 0 is physically irrelevant. Thus
the problem of objects entering or leaving the spacetime along
some null surface has been solved.

To summarize: the Poincar$\m{\acute{e}}$ ``disc" model of H$^4$ can
be complexified, but \emph{only} in the
($+\;+\;+\;+$)$\rightarrow$($+\;-\;-\;-$) sense. The result is the
very interesting Spatially Hyperbolic version of de Sitter
spacetime.

Finally, we turn to the foliation of hyperbolic space which is
actually the one used by Ooguri et al \cite{kn:OVV}.

\subsubsection*{{\textsf{3.4. Foliation Corresponding to Spatially Flat de Sitter}}}
We define coordinates $\Phi$, x, y, z on H$^4$ by
\begin{eqnarray} \label{eq:FLANEUR}
\m{A} & = & \m{L\;cosh(\Phi/L)\;+\;{{1}\over{2L}}\,(x^2\;+\;y^2\;+\;z^2)\,e^{-\,\Phi/L}  }     \nonumber \\
\m{B} & = & \m{L\;sinh(\Phi/L)\;+\;{{1}\over{2L}}\,(x^2\;+\;y^2\;+\;z^2)\,e^{-\,\Phi/L}}    \nonumber \\
\m{Z} & = & \m{z\,e^{-\,\Phi/L}}           \nonumber \\
\m{Y} & = & \m{y\,e^{-\,\Phi/L}}  \nonumber \\
\m{X} & = & \m{x\,e^{-\,\Phi/L}}.
\end{eqnarray}
Here all coordinates run from $-\,\infty$ to $+\,\infty$, and the
metric is
\begin{equation}\label{eq:HORREUR}
g(\m{H^4; \Phi)_{++++} \;=\; d\Phi^2\;
+\;e^{(-\,2\,\Phi/L)}\,[\,dx^2 \;+\; dy^2 \;+\; dz^2]}.
\end{equation}
Evidently the surfaces $\Phi$ = constant are infinite, flat spaces
of topology $\bbr^3$. In order to understand how these fit into
the Poincar$\m{\acute{e}}$ disc, recall that, by stereographic
projection, $\bbr^3$ has the same topology as a three-sphere from
which one point has been deleted. If, therefore, we take a finite
sphere in the Poincar$\m{\acute{e}}$ disc and move it until it
touches the boundary sphere at one point [which we take to be the
``north pole" of the disc coordinates, $\chi$ = 0 in equations
(\ref{eq:HYPER})], the part of the sphere which lies in the bulk
is in fact a copy of $\bbr^3$ [see Figure 4]. A collection of such
copies of $\bbr^3$, all obtained from spheres intersecting at the
same point on the conformal boundary, foliate the entire bulk of
H$^4$; see Figure 4. The metric corresponding to this foliation is
precisely the one given in (\ref{eq:HORREUR}). The value of $\Phi$
corresponds to the size of the sphere in the figure: larger
spheres correspond to negative values of $\Phi$, which runs from
bottom to top in the figure.
\begin{figure}[!h]
\centering
\includegraphics[width=0.8\textwidth]{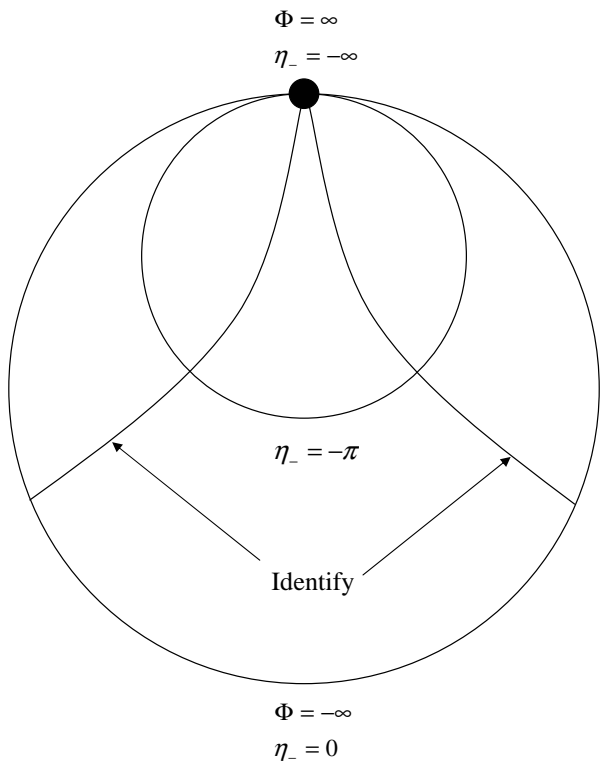}
\caption{$\bbr^3$ foliation of H$^4$, and its partial
compactification.}
\end{figure}

As in the cases with hyperbolic spatial sections, the Lorentzian
version of this geometry will be geodesically incomplete along a
null surface, ruling out ``creation from nothing", unless we
compactify the $\bbr^3$ slices. They can of course be compactified
to topology $\bbr^3/\bbz^3$ [among other possibilities] by simply
setting x = K$\theta_1$, y = K$\theta_2$, z = K$\theta_3$, for
some positive constant K, where $\theta_{1,2,3}$ are angular
coordinates. The slices are now cubic tori, with a size which
``evolves" from infinity at $\Phi$ = $-\,\infty$ to zero at $\Phi$
= $+\,\infty$. The metric is now
\begin{equation}\label{eq:HANNIBAL}
g(\m{H}^4/\bbz^3; \Phi)_{++++} \;=\; \m{d\Phi^2\;
+\;K^2\,e^{(-\,2\,\Phi/L)}\,[\,d\theta_1^2 \;+\; d\theta_2^2 \;+\;
d\theta_3^2]}.
\end{equation}
This is just the four-dimensional version of the metric (\ref{eq:P})
considered by Ooguri et al \cite{kn:OVV}. The manifold can be
approximately portrayed as the region between the lines extending
down from the north pole in Figure 4: the idea is that one keeps
only a finite piece of each $\bbr^3$ section of the full space, and
then performs the required identifications on this piece.

This space has two complexifications. One can of course complexify
$\theta_1$, thereby obtaining [``Partly Compactified"] anti-de
Sitter spacetime:
\begin{equation}\label{eq:HANNIBAL2}
g(\m{PCAdS_4})_{-+++} \;=\; \m{d\Phi^2\;
+\;K^2\,e^{(-\,2\,\Phi/L)}\,[\,-\;d\theta_1^2 \;+\; d\theta_2^2
\;+\; d\theta_3^2]}.
\end{equation}
But there is another, less obvious complexification.

Ooguri et al observe that one cannot complexify $\Phi$ here, so
that one cannot obtain a complexification of the
($+\;+\;+\;+$)$\rightarrow$($-\;+\;+\;+$) kind in that way.
Instead we proceed in the now familiar manner: we search for a
complexification of the form
($+\;+\;+\;+$)$\rightarrow$($+\;-\;-\;-$). This can be achieved
here in a particularly elegant manner if we define a dimensionless
[angular] Euclidean ``conformal time" $\eta_-$, taking values in
the range ($-\,\infty$, 0), by
\begin{equation}\label{eq:HAMMERKLAVIER}
\eta_-\;=\;-\,\pi\,\m{e^{\Phi/L}}.
\end{equation}
We now have
\begin{equation}\label{eq:HARGITTAI}
g(\m{H}^4/\bbz^3;\eta_-)_{++++} \;=\;\m{{{1}\over{\eta_-^2}}\,[\,
L^2d\eta_-^2\; +\;\pi^2\,K^2\,\{d\theta_1^2 \;+\; d\theta_2^2
\;+\; d\theta_3^2\}]}.
\end{equation}
We now complexify $\eta_-\,\rightarrow\,\pm\m{i}\eta_+$, where
$\eta_+$ takes its values in (0, $\infty$). The result is the
well-known spatially flat version of Lorentzian de Sitter spacetime,
but now with toral sections: it is Spatially Toral de Sitter, in
($+\;-\;-\;-$) signature:
\begin{eqnarray}\label{eq:HAMMURABI}
g(\m{STdS_4})_{+---} & = & \m{{{1}\over{\eta_+^2}}\,[\,
L^2d\eta_+^2\; -\;\pi^2\,K^2\,\{d\theta_1^2 \;+\; d\theta_2^2
\;+\; d\theta_3^2\}]} \nonumber \\
                     & = &  \m{dt^2\; -\;K^2\,e^{(2\,t/L)}\,[\,d\theta_1^2 \;+\;
d\theta_2^2 \;+\; d\theta_3^2]},
\end{eqnarray}
where t, which ranges from $-\,\infty$ to $+\,\infty$, is related to
$\eta_+$ by
\begin{equation}\label{eq:HORATIO}
\eta_+\;=\;+\,\pi\,\m{e^{-\,t/L}}.
\end{equation}

We stress again that this spacetime is topologically and physically
\emph{distinct} from both the Spatially Spherical and the Spatially
Hyperbolic de Sitter spacetimes.

Thus we have succeeded in associating an accelerating Lorentzian
cosmology with the partially compactified hyperbolic space
H$^4$/$\bbz^3$, just as Ooguri et al require.

In Section 2 we saw that there were two distinct complexifications
of the sphere, arising from two distinct but equally valid ways of
foliating it. We also saw, however, that one complexification was
favoured over the other, because one of the resulting Lorentzian
spacetimes failed to attain macroscopic size: it collapsed
immediately after being created. In the hyperbolic case, by
contrast, nothing of this sort happens: one complexification leads
to AdS$_4$, while the other three lead to three physically
distinct families\footnote{``Families", because in each case there
are many possible compactifications of the spatial sections: see
\cite{kn:rp3}\cite{kn:reallyflat}\cite{kn:silver1}.} of expanding,
accelerating spacetimes. How can this ambiguity be resolved?

In order to answer this, let us consider the peculiarities of the
two-dimensional case, since that is the case discussed in
\cite{kn:OVV}.

\addtocounter{section}{1}
\section*{\large{\textsf{4. The Two-Dimensional Case and What it Teaches Us }}}

In the two-dimensional case, Figures 1, 2, 3, and 4 can be
interpreted literally, in the sense that there are no angles to be
suppressed. In particular, a simple reflection,
$\Psi\,\rightarrow\,\rho$, $\Sigma\,\rightarrow\,\Theta$ shows that
the two foliations in Figures 1 and 2 are identical. More
interesting is the fact that the corresponding Lorentzian spaces are
also identical: in the two-dimensional case we have [from equations
(\ref{eq:L}) and (\ref{eq:M})]
\begin{equation}\label{eq:O}
g(\m{AdS_2)_{-\,+}} \;=\; \m{-\,cosh^2(S/L) \, dU^2\;+ \;dS^2},
\end{equation}
\begin{equation}\label{eq:OCELOT}
g\m{(SSdS_2)_{+\,-}\;=\; dT^2\;-\;L^2\,cosh^2(T/L)\,d\chi^2},
\end{equation}
where, as before, we regard both U/L and $\chi$ as angular
coordinates. This corresponds to the fact that the anti-de Sitter
group in n+1 dimensions, O(2,n), is isomorphic to the de Sitter
group O(n+1,1), when n = 1. [Actually the symmetry groups are
smaller, since we compactify U and $\chi$, but these smaller groups
are the same for both spacetimes]. It is helpful to consider how
``Euclidean time" works in Figure 3. In the ``AdS" case one thinks
of ``time" as running vertically [so that the boundary is at
``spatial" infinity], while in the ``dS" case one reflects the
diagram about a diagonal so that the boundary is in the ``future"
and ``past". Of course, neither definition of Euclidean ``time" is
more valid than the other, and this statement is the basis of the
dual interpretation of two-dimensional hyperbolic space introduced
by Ooguri et al.

In view of this, the claim of Ooguri et al, that one can create
from ``nothing" an accelerating two-dimensional cosmology using [a
version of] \emph{negatively} curved hyperbolic space, clearly
must be valid. All that remains is to see how the derivation works
in a technical sense. We claim that the necessary technical device
is precisely the very mild generalization of complexification that
we have introduced here. Let us see how this works in detail.

First, take equation (\ref{eq:OCELOT}) and split the spacetime
along its spacelike hypersurface of zero extrinsic curvature at T
= 0, retaining only the T $\geq$ 0 half. Similarly we can take the
two-dimensional version of equation (\ref{eq:K}), compactify the
coordinate $\rho$ [so that we are dealing with H$^2$/$\bbz$ rather
than H$^2$ itself], and obtain
\begin{eqnarray}\label{eq:KAOS}
g(\mathrm{H}^2/\bbz; \Theta)_{++}\; =\;\m{L^2}\,
[\m{d\Theta^2}\;+\;\m{cosh^2(\Theta)}\mathrm{d\rho^2}];
\end{eqnarray}
this too splits naturally at $\Theta$ = 0, and we retain only
$\Theta\,\leq\,$ 0. Both T = 0 and $\Theta$ = 0 are surfaces of zero
extrinsic curvature, and both are circles of radius L. We combine
the two halves along these surfaces, obtaining a manifold of
topology $\bbr\,\times\,\m{S}^1$ with a metric which we symbolize by
\begin{eqnarray}\label{eq:KURSE}
g(\mathrm{H}^2/\bbz; \Theta\,\leq\,0)_{++}\;
\longrightarrow\;g\m{(SSdS_2;\,T\,\geq\,0)_{+\,-}}.
\end{eqnarray}
[The arrow here represents the idea that the Euclidean version is
succeeded by the Lorentzian version at the creation.] Of course,
this just means that we have a Euclidean-to-Lorentzian transition,
with the first metric valid on one side, the second on the other.
This would be the geometry underlying the creation of this version
of two-dimensional de Sitter spacetime, as described by the Ooguri
et al negatively-curved analogue of the hemisphere used to construct
the Hartle-Hawking wave function.

We can repeat this for the other two versions of two-dimensional de
Sitter: from equation (\ref{eq:OLIPHAUNT}) we have
\begin{eqnarray}\label{eq:KAKOPHONY}
g(\mathrm{SHdS_2})_{+\,-}\; =\;+\;
\m{d\tau}^2\;-\;\m{sinh^2(\tau/L)}\,\mathrm{dr^2},
\end{eqnarray}
where our agreed compactification of the spatial sections means
that r is proportional to some angular coordinate, so that the
$\tau$ = constant sections are circular. This is to be compared
with the two-dimensional version of (\ref{eq:ORNITHOPTER}),
\begin{equation}\label{eq:KORRAL}
g(\mathrm{H^2; P})_{++}\; =\;+\;
\m{dP}^2\;+\;\m{L^2}\,\m{sinh^2(P/L)\,d\chi^2;}
\end{equation}
here it will be useful to define P as having negative values
ranging from $-\,\infty$ to zero.

Notice that there is an important difference between this case and
the previous one: here there is no surface of zero extrinsic
curvature in either the Lorentzian or the Euclidean cases. As this
is also a property of the spatially flat case considered by Ooguri
et al [see below], we postpone discussion of this point; for the
moment let us arbitrarily truncate the range of $\tau$ to
[$\alpha$, $\infty$) for some positive constant $\alpha$ with
dimensions of length, and that of P to ($-\,\infty$, $-\,\alpha$],
so that at both $\tau$ = $\alpha$ and P = $-\,\alpha$ we have
circular sections of radius Lsinh($\alpha$/L). Joining the two
spaces along these circles, we obtain a space of topology
$\bbr\,\times\,\m{S}^1$ with a metric
\begin{eqnarray}\label{eq:KALLIGRAPHY}
g(\mathrm{H^2; P\,\leq\,-\,\alpha})_{++}\;
\longrightarrow\;g\m{(SHdS_2;\,\tau\,\geq\,\alpha)_{+\,-}}.
\end{eqnarray}
Assuming that the truncations can be justified, this would be the
geometry underlying the description by the OVV wave function of the
creation of Spatially Hyperbolic de Sitter spacetime from
``nothing".

Finally we come to the case actually studied by Ooguri et al, with
flat, toral spatial sections. The Euclidean metric in this case is
just the two-dimensional version of (\ref{eq:HANNIBAL}) and
(\ref{eq:HARGITTAI}),
\begin{eqnarray}\label{eq:K1}
g(\m{H}^2/\bbz; \eta_- ; \Phi)_{++}
\;&=&\;\m{{{1}\over{\eta_-^2}}\,[\,L^2\,d\eta_-^2\;
+\;\pi^2\,K^2\,d\theta_1^2]}
\nonumber\\
\;&=& \;\m{d\Phi^2\; +\;K^2\,e^{(-\,2\,\Phi/L)}\,d\theta_1^2},
\end{eqnarray}
and of course we wish to combine this with the two-dimensional
version of (\ref{eq:HAMMURABI}),
\begin{eqnarray}\label{eq:K2}
g(\m{STdS_2})_{+-} & = &
\m{{{1}\over{\eta_+^2}}\,[\,L^2\,d\eta_+^2\;
-\;\pi^2\,K^2\,d\theta_1^2}] \nonumber \\
                     & = &  \m{dt^2\;
                     -\;K^2\,e^{(2\,t/L)}\,d\theta_1^2}.
\end{eqnarray}
As in the case of hyperbolic sections, the absence of any surface
of zero extrinsic curvature here means that we have to truncate
the ranges of $\eta_+$ and $\eta_-$. Since the other coordinate,
$\theta_1$, is angular [ranging from $-\,\pi$ to $+\,\pi$], it is
natural to truncate $\eta_-$ at $\eta_-$ = $-\,\pi$, so that the
range of this coordinate is [$-\,\pi$, 0) --- see Figure 4. This
truncates the space along a circle which [by equation
(\ref{eq:K1})] is of circumference 2$\pi$K. To ensure continuity,
the Lorentzian spacetime must also be truncated along a circle of
circumference 2$\pi$K. This, by equation (\ref{eq:K2}), means that
the range of Lorentzian conformal time is (0, $\pi$]. We then
topologically identify the circle at $\eta_-$ = $-\,\pi$ with the
circle at $\eta_+$ = $\pi$; \emph{thus, K is the initial radius of
the Universe at the moment of creation}.

The fact that we are taking the range of all ``angular" coordinates
to be from $-\,\pi$ to $+\,\pi$ now neatly reflects the fact that
the Euclidean and Lorentzian spaces are identified along their
edges. Note that this angular interpretation of Euclidean and
Lorentzian conformal time suggests that we should consider a
topological identification of the Euclidean conformal boundary
[$\eta_-$ = 0] with the Lorentzian future spacelike conformal
infinity [$\eta_+$ = 0]. The full conformal compactification will
then itself be a [two-dimensional] torus. Roughly speaking, L is the
radius of this torus in one direction, while K is its radius in the
other. We shall return to this in the Conclusion.

We now define a metric on $\bbr\,\times\,\m{S}^1$ by
\begin{eqnarray}\label{eq:K3}
g(\m{H^2}/\bbz;\; -\,\pi\,\leq\,\eta_-\,<\,0)_{++}\;
\longrightarrow\;g(\m{STdS_2;\;0\,<\, \eta_+\,\leq\,\pi})_{+-}.
\end{eqnarray}
\emph{This is the geometry describing the creation of an
accelerating Universe, with flat [but compact] spatial sections, in
the OVV picture.}

The fact that one has to truncate both the spatially hyperbolic and
the spatially toral versions of de Sitter spacetime is due to the
structure of the Einstein equations [in the form of the Friedmann
equations], which forbid the extrinsic curvature of any hypersurface
to vanish. This means that the Euclidean-to-Lorentzian transition
can be continuous but not smooth. Of course, one has every reason to
suspect that the Einstein equations do not hold exactly at the
transition point, and that the corrected equations will allow a
smooth transition. This proves to be so, and we shall discuss the
details [for toral sections] when we return to the four-dimensional
case in the next section.

The two-dimensional case is interesting partly because it arises
naturally in the context considered by Ooguri et al, and partly
because it teaches us that there must be some natural way of
associating an accelerating Lorentzian universe with a negatively
curved Euclidean space. We have argued that there is indeed a very
simple way of establishing such an association: complexify according
to [the two-dimensional version of]
($+\;+\;+\;+$)$\rightarrow$($+\;-\;-\;-$) instead of
($+\;+\;+\;+$)$\rightarrow$($-\;+\;+\;+$). If we proceed in this
way, we find that we can set up the geometric background for the OVV
version of creation from ``nothing": indeed, we can do this for all
three versions of de Sitter spacetime. This last point is somewhat
disappointing, since one might have hoped that the OVV wave function
might give us a clue as to which version is the correct one. As we
shall now see, the situation in four dimensions is much more
satisfactory in this regard.

\addtocounter{section}{1}
\section*{\large{\textsf{5. Four Dimensions}}}
In four dimensions, we are again interested in three metrics of
constant negative curvature: $g(\mathrm{H}^4; \Theta)_{++++}$
[equation (\ref{eq:K})], $g(\mathrm{H^4; P})_{++++}$ [equation
(\ref{eq:ORNITHOPTER})], and $g(\m{H}^4/\bbz^3)_{++++}$ [equation
(\ref{eq:HARGITTAI})].

But now we find something remarkable: if we try to generalize the
discussion of the preceding section to the four-dimensional case, it
\emph{does not work} for $g(\mathrm{H}^4; \Theta)_{++++}$ and
$g(\mathrm{H^4; P})_{++++}$. For the transverse sections defined by
$g(\mathrm{H}^4; \Theta)_{++++}$ are negatively curved: they cannot
be joined to the spacelike sections of the Lorentzian version,
$g(\m{SSdS_4)_{+---}}$ [equation (\ref{eq:M})], since these are
positively curved. Similarly, the transverse sections defined by
$g(\mathrm{H^4; P})_{++++}$ are positively curved, and cannot be
joined to the negatively curved spacelike sections of
$g(\mathrm{SHdS_4})_{+\,---}$ [equation (\ref{eq:OLIPHAUNT})].

This difficulty did not arise in the two-dimensional case, for the
simple reason that the spatial sections of a two-dimensional
spacetime are one-dimensional, and of course one-dimensional
manifolds cannot be curved either positively or negatively. The
only case where this is not a problem in higher dimensions is the
case where the spatial sections are \emph{flat}, since a reversal
of the sign of the curvature has no effect here. This is of course
precisely the case considered by Ooguri et al.

Hartle \cite{kn:hartle} has recently argued that Lorentzian
signature ``emerges" from the formalism of Euclidean quantum
gravity. If we begin with a spherical Euclidean geometry, the
geometry cannot remain Euclidean if its sections are to become
significantly larger than the curvature scale: there has to be a
transition to a different signature for this to be possible. But
this signature cannot be to ($+\;+\;-\;-$) or ($-\;-\;+\;+$), since
the three-dimensional sections in such a case would themselves be
Lorentzian, not Euclidean. Thus Hartle claims that Lorentzian
signature is ``emergent" within this theory of quantum gravity. In
precisely the same way, we find that \emph{the toral structure of
the spatial sections of our Universe is emergent within the theory
of Ooguri et al:} foliations of the Euclidean space with either
positively or negatively curved sections are unable to make the
transition to the Lorentzian regime\footnote{Of course, a
three-dimensional torus with signature ($-\;-\;-$) has essentially
the same geometry as a torus with signature ($+\;+\;+$).}.

In the toral case, and in this case \emph{only}, we can generalize
the discussion of the preceding section: we truncate
H$^4$/$\bbz^3$ so that $\eta_-$ takes values in [$-\,\pi$, 0),
while Spatially Toral de Sitter, STdS$_4$, is truncated so that
$\eta_+$ takes values in (0, $\pi$]; both spaces are joined along
a three-dimensional torus consisting of circles of circumference
2L/$\beta$. Then we can define a metric on $\bbr\,\times\,\m{T}^3$
by
\begin{eqnarray}\label{eq:K6}
g(\m{H^4}/\bbz^3;\; -\,\pi\,\leq\,\eta_-\,<\,0)_{++++}\;
\longrightarrow\;g(\m{STdS_4;\;0\,<\, \eta_+\,\leq\,\pi})_{+---},
\end{eqnarray}
where the change of signature is effected by complexifying
\emph{conformal} time. [Recall that the arrow symbolizes the
transition from a Euclidean to a Lorentzian metric, and bear in
mind that the Euclidean-to-Lorentzian transition occurs along
$\eta_-$ = $-\,\pi$ and $\eta_+$ = $\pi$, not at $\eta_{\pm}$ =
0.] This should describe the creation of the universe [at $\eta_-$
= $-\,\pi$ in Figure 4] in terms of a four-dimensional version of
the Ooguri et al wave function, defined on H$^4$/$\bbz^3$. Again,
it may be of interest to consider identifying the Euclidean
infinity at $\eta_-$ = 0 with Lorentzian future spacelike infinity
at $\eta_+$ = 0, as suggested by the angular interpretation of
$\eta_+$ and $\eta_-$.

As in the two-dimensional case, the Euclidean-to-Lorentzian
transition here is continuous but not smooth; this must be resolved
by a suitable modification of the Einstein equations in that region
of spacetime. Notice that while this was optional in the
two-dimensional case, it is compulsory here, since we have argued
that the sections \emph{must} be toral; FRW models with toral
sections cannot have a section of zero extrinsic curvature if only
non-exotic matter is present and if the Einstein equations hold
exactly.

A concrete suggestion as to how the smoothing occurs was made in
\cite{kn:singularstable}, where it was proposed that the structure
responsible was the ``classical constraint field" proposed by
Gabadadze and Shang \cite{kn:gab1}\cite{kn:gab2}. This leads, in
the four-dimensional case, to a Friedmann equation of the form
\begin{equation}\label{eq:GABADAD}
\m{\Big({{\dot{a}\over{a}}\Big)^2
\;=\;{{1}\over{L_{\m{inf}}^2}}\;-\;{{b\,\varepsilon}\over{6\,a^6}}\,}},
\end{equation}
where we assume that the ``constraint field" is significant during a
short interval between the creation of the Universe and a subsequent
inflationary era characterized by a length scale L$_{\m{inf}}$, and
where b$\varepsilon$ is a certain constant which may in general be
positive or negative\footnote{In the model of Gabadadze and Shang,
the spatial sections are flat manifolds-with-boundary, but the idea
of the constraint field also works for flat compact sections,
assumed here.}. If we wish to create such a universe from
``nothing", however, b$\varepsilon$ is fixed by the requirement that
the Euclidean/Lorentzian transition be smooth: this imposes
\begin{equation}\label{eq:K4}
\m{b\,\varepsilon\;=\;6/L_{inf}^2},
\end{equation}
where we take it that the scale function is equal to unity at the
transition. Substituting this into equation (\ref{eq:GABADAD}), we
obtain an equation which can be solved exactly; the resulting metric
is, in the notation of \cite{kn:singularstable},
\begin{equation}\label{eq:K5}
g\m{_c(6,\,K,\,L_{inf})_{+---} \;=\; dt^2\; -\;
K^2\;cosh^{(2/3)}\Big({{3\,t}\over{L_{inf}}}\Big)\,[d\theta_1^2
\;+\; d\theta_2^2 \;+\; d\theta_3^2]};
\end{equation}
here K is the parameter which fixes the size of the initial torus.
Notice that, as t tends to positive infinity, this metric quickly
becomes indistinguishable from that of Spatially Toral de Sitter
spacetime, $g(\m{STdS_4})_{+---}$. Thus, the spacetime has a
spacelike future conformal infinity.

This metric can be expressed in terms of a dimensionless
Lorentzian conformal time $\eta_+$, defined by
\begin{equation}\label{eq:AUDI1}
\m{\eta_+\;=\;{{\beta}\over{3}}\,\int_{3t/L_{inf}}^{\infty}sech^{1/3}(x)\,dx},
\end{equation}
where $\beta$ is the constant defined by
\begin{equation}\label{eq:AUDI4}
\beta
\;=\;\m{{{3\,\pi}\over{\int_0^{\infty}sech^{1/3}(x)\,dx}}\;\approx\;2.5871
}.
\end{equation}
Notice that the integral in equation (\ref{eq:AUDI1}) will
converge even if proper time is integrated to infinity. This means
that the smoothing of the geometry at the transition point
automatically truncates the conformal time to a finite range ---
we do not have to do this by hand, as we did in the case of pure
Spatially Toral de Sitter spacetime. In view of this, we have
defined $\eta_+$ so that, as in the case of STdS$_4$ discussed
earlier, it is equal to zero at t = $\infty$, and $\eta_+$ = $\pi$
at t = 0; thus $\eta_+$ ranges from 0 to $\pi$ [though this way of
putting things reverses the usual direction of time].

If one were able to do the integration\footnote{It can of course
be done, in terms of hypergeometric functions; the result is not
useful, however.} in equation (\ref{eq:AUDI1}), one would be able
to express t/L$_{\m{inf}}$ as a function of $\eta_+$ and so the
scale function can likewise be regarded as a function of $\eta_+$.
Let us define a function G($\eta_+$) by
\begin{equation}\label{eq:AUDI2}
\beta\,\m{G(\eta_+)\;=\;cosh^{1/3}(3t/L_{inf})}.
\end{equation}
Notice that this implies that G($\pi$) = 1/$\beta$. Since the
right side of this equation is approximated by
(e$^{\m{t/L_{inf}}}$)/2$^{(1/3)}$ when t is large, it follows from
equation (\ref{eq:HORATIO}) that for $\eta_+$ close to zero, we
have
\begin{equation}\label{eq:AUDITT}
\beta\,\m{G(\eta_+)\;\approx \;{{\pi}\over{2^{1/3}\,\eta_+}}}.
\end{equation}
This relation implies that complexifying $\eta_+$ necessarily
entails complexifying G($\eta_+$); that is,
$\eta_+\,\rightarrow\,\pm\m{i}\eta_-$ implies
$\m{G(\eta_+)\,\rightarrow\,\mp iG(\eta_-)}$, where $\eta_-$ is
Euclidean conformal time, which takes its values in [$-\,\pi$ , 0),
with the transition at $-\,\pi$ and Euclidean infinity at $\eta_-$ =
0.

It follows that if we write $g\m{_c(6,\,K,\,L_{inf})_{+---}}$ in the
form
\begin{equation}\label{eq:AUDI3}
g\m{_c(6,\,K,\,L_{inf})_{+---} \;=\; G(\eta_+)^2\,[\,
L_{inf}^2d\eta_+^2\; -\;\beta^2\,K^2\,\{d\theta_1^2 \;+\;
d\theta_2^2 \;+\; d\theta_3^2\}]},
\end{equation}
then its Euclidean version is just
\begin{equation}\label{eq:BMW1}
g\m{_c(6,\,K,\,L_{inf})_{++++} \;=\; G(\eta_-)^2\,[\,
L_{inf}^2d\eta_-^2\; +\;\beta^2\,K^2\,\{d\theta_1^2 \;+\;
d\theta_2^2 \;+\; d\theta_3^2\}]}.
\end{equation}
This can of course be written as
\begin{equation}\label{eq:K7}
g\m{_c(6,\,K,\,L_{inf})_{++++} \;=\; d\Phi^2\;
+\;K^2\;cosh^{(2/3)}\Big({{3\,\Phi}\over{L_{inf}}}\Big)\,[d\theta_1^2
\;+\; d\theta_2^2 \;+\; d\theta_3^2]},
\end{equation}
where $\Phi$ runs from $-\,\infty$ to zero. This is
indistinguishable from $g(\m{H}^4/\bbz^3)_{++++}$ for sufficiently
large negative $\Phi$. Thus $g\m{_c(6,\,K,\,L_{inf})_{++++}}$ and
$g\m{_c(6,\,K,\,L_{inf})_{+---}}$ interpolate between the two
metrics we wish to join, \emph{but this is done smoothly}.

The full metric, combining the Euclidean and Lorentzian versions of
the metric on a manifold of topology $\bbr\,\times\,\m{T}^3$, is now
\begin{equation}\label{eq:K8}
g\m{_c(6,\,K,\,L_{inf};
-\,\pi\,\leq\,\eta_-\,<\,0)_{++++}}\;\longrightarrow\;g\m{_c(6,\,K,\,L_{inf};0\,<\,\eta_+\,\leq\,\pi)_{+---}}.
\end{equation}
The Euclidean-to-Lorentzian transition is at $\eta_-$ = $-\,\pi$ and
$\eta_+$ = $\pi$, that is, at t = $\Phi$ = 0. The two spaces are
joined along a torus of radius K in a way such that the scale factor
of the full metric is infinitely differentiable.

Note that the angular interpretation of conformal time seems
particularly natural in this case, since no truncations have to be
performed by hand. If one identifies $\eta_+$ = 0 with $\eta_-$ = 0,
then the conformal compactification is fully toral: it is a
four-torus. The inflationary length scale L$_{\m{inf}}$ determines
the radius of one circle, while K determines the radius of the other
three. [Strictly speaking, in conformal geometry only the ratio
K/L$_{\m{inf}}$ is a well-defined parameter here, so one should say
that the initial size of the Universe measured in inflationary units
is what fixes the shape of the conformal torus.]

Of course, the suggestion that the Gabadadze-Shang constraint field
is responsible for the smoothing is just one possibility; there are
others; the main point is that the smoothing can be done in a
physical way. It seems reasonable, however, to assert that the
metric given in (\ref{eq:K8}) is the simplest possible smooth model
of a four-dimensional Euclidean space of the OVV type giving rise to
an accelerating Lorentzian cosmological model.

\addtocounter{section}{1}
\section*{\large{\textsf{6. Conclusion}}}
We can summarize as follows. The OVV wave function is formulated
on [a partially compactified version of] hyperbolic space,
H$^2/\bbz$. By means of a simple extension of the concept of
complexification, we have been able to explain how to realize the
idea of Ooguri et al that this space has two Lorentzian
interpretations, one [equation (\ref{eq:HANNIBAL2})] like anti-de
Sitter, the other [equation (\ref{eq:HAMMURABI})] like de Sitter
spacetime. This idea works in all dimensions, but, in dimensions
above two, it \emph{only} works in the case where the transverse
sections are flat tori. The global structure of the
three-dimensional sections of our Universe is thus
\emph{emergent}, in Hartle's \cite{kn:hartle} sense, from the OVV
formalism.

The flatness of the spatial sections requires that Einstein's
equations be corrected near the creation event, so that the
Euclidean/Lorentzian transition can be smooth; we have suggested a
concrete way, based on the ideas of Gabadadze and Shang
\cite{kn:gab1}\cite{kn:gab2}, whereby the smoothing can be achieved
in a physical manner.

Obviously a great deal remains to be done. First, one must indeed
extend the OVV theory to four dimensions. The first step would be
to replace H$^2/\bbz$ with its natural higher-dimensional version,
the partial compactification H$^4/\bbz^3$. Ooguri et al embed
H$^2/\bbz$ in IIB string theory by considering a background of the
form $(\m{H^2}/\bbz)\,\times\,\m{S^2\,\times\,CY}$, where CY
denotes some Calabi-Yau manifold. One way to embed H$^4/\bbz^3$ in
string theory
--- or, rather, M-theory --- might be through a compactification of the form
$(\m{H^4}/\bbz^3)\,\times\,\m{FR}$, where FR denotes a [singular]
Freund-Rubin space of the kind studied in \cite{kn:bobby}.
Extending the OVV ideas to spaces of this kind is a challenging
problem. If it can be done, the next step would be to try to
understand the consequences of smoothing the Euclidean/Lorentzian
transition in this context. One would then be able, by means of a
complexification of the kind suggested here, to see what the OVV
theory predicts regarding the nature of four-dimensional
accelerating cosmologies.

We argued above that it is natural to think of $\eta_+$ and $\eta_-$
as angular variables: the identification of the two spaces at the
Euclidean-to-Lorentzian point is then expressed by the familiar fact
that $-\,\pi$ and $+\,\pi$ refer to the same point in plane polar
coordinates. If we take this to its logical conclusion, then we
should \emph{also} identify $\eta_-$ = 0 with $\eta_+$ = 0. In other
words, Euclidean conformal infinity is just Lorentzian future
spacelike infinity, approached from the ``other side". In this case
the full conformal compactification of the combined space with
metric (\ref{eq:K6}) or (\ref{eq:K8}) has the topology of a
four-dimensional torus. This way of thinking could possibly be of
interest in connection with ideas about holography at future
spacelike infinity in accelerating cosmologies\footnote{See in
particular the discussion of this kind of holography in
\cite{kn:polchinski}.}. In particular, it might help us to
understand how a necessarily Euclidean conformal field theory at
future spacelike infinity can be dual to Lorentzian physics. The
idea would be that the duality is effective through the
Euclidean-to-Lorentzian transition, by going the ``long way" around
the circle, clockwise from $\eta_-$ = 0 to $\eta_-$ = $-\,\pi$,
through the Euclidean-to-Lorentzian transition there, from which the
Universe evolves in Lorentzian conformal time from $\eta_+$ = $\pi$
back to future spacelike infinity at $\eta_+$ = 0.

The minisuperspace construction considered by Ooguri et al leads
to a prediction that the most probable geometry is flat space;
this is reminiscent of the conclusion, derived from the
Hartle-Hawking wave function, that the most probable value of the
cosmological constant is zero \cite{kn:zero}. In the latter case
it has been argued convincingly in
\cite{kn:tye}\cite{kn:sarangi}\cite{kn:sashtye} that the wave
function needs to be modified in some way that will involve going
beyond the most basic minisuperspace constructions. We have seen
here that, in the OVV case, one has to go beyond the most basic
minisuperspace models merely to obtain a smooth
Euclidean/Lorentzian transition, leading to a metric which could
resemble the one given in (\ref{eq:K8}) above. Perhaps this
geometry will be useful in an attempt to extend the ideas of
\cite{kn:tye}\cite{kn:sarangi}\cite{kn:sashtye} to the OVV wave
function, so that more reasonable predictions can be made. The
obvious first step would be to try to predict the value of the
parameter K in (\ref{eq:K8}), to see whether the modified wave
function predicts a physically acceptable value for the initial
size of the Universe. A value close to the string length scale
would be particularly interesting: for that would implement
T-duality for the circles constituting the initial three-torus, in
the sense that no circles of radius smaller than the string scale
would ever exist.

 \addtocounter{section}{1}
\section*{\textsf{Acknowledgements}}
The author is extremely grateful to Wanmei for preparing the
diagrams.

\end{document}